\documentclass[a4paper,12pt]{article}
\usepackage{graphicx,amsmath,amsfonts,amssymb,cite}
\newcommand{\be}{\begin{equation}}
\newcommand{\ee}{\end{equation}}

\begin{document}

\begin{center}
{\Large\bf Weinberg like sum rules revisited}
\end{center}

\begin{center}
{\bf S. S. Afonin\footnote{E-mail: afonin24@mail.ru,
afonin29@yandex.ru}}
\end{center}

\begin{center}
{\it V. A. Fock Department of Theoretical Physics, St. Petersburg
State University, 1 ul. Ulyanovskaya, 198504, St. Petersburg,
Russia.}
\end{center}

\begin{abstract}
The generalized Weinberg sum rules containing the difference of
isovector vector and axial-vector spectral functions saturated by
both finite and infinite number of narrow resonances are considered.
We summarize the status of these sum rules and
analyze their overall agreement with phenomenological
Lagrangians, low-energy relations, parity doubling, hadron string
models, and experimental data.
\end{abstract}

\section{Introduction}

The sum rules equating certain moments of the spectral weight
functions of two-point current correlators turned out to be an
extremely fruitful concept in the hadron physics. They were first
proposed by Weinberg who considered the difference of vector ($V$)
and axial-vector ($A$) correlators~\cite{wein}, later this difference
was recognized to be an order parameter of spontaneous chiral
symmetry breaking in QCD in the chiral limit.
The original derivations of such sum
rules suffered from a lack of rigour as long as they were based on
{\it ad hoc} postulates about the high-energy behaviour of certain
combinations of two-point correlators and about the possible
nature of Schwinger terms of the current commutators. The original
Weinberg's derivation rested on a proof of equality of $V$ and $A$
Schwinger terms and on the assumption of asymptotic chiral $SU(2)\times SU(2)$
symmetry at high momentum, thus providing for the first time a
concrete realization of the notion of chiral symmetry at short
distances. The latter proposal was immediately generalized
in~\cite{das2}, where it was shown that the idea of asymptotic
symmetries may serve as a powerful tool for deriving various
interesting results in the pole approximation. These papers were
followed by the $SU(3)\times SU(3)$ generalization of Weinberg sum
rules~\cite{wein2} and subsequent early applications to hadron
physics~\cite{wein3}. The Weinberg sum rules were exploited for
derivation of electromagnetic mass difference~\cite{das} and
Das-Mathur-Okubo sum rule~\cite{das3}, both results have been
widely used up to now. The former one relies essentially on the
validity of the second Weinberg sum rule, otherwise this mass
difference is not finite\footnote{Later it was
demonstrated~\cite{dicus} that the use of the second
spectral-function sum rule is not necessary if an exchange by
intermediate weak gauge boson is taken into account.}. This and
other phenomenological observations called for a rigorous
justification of spectral-function sum rules. An important
requirement was also the universality of such a justification, as,
say, the original method used in~\cite{wein} to derive the
equality of $V$ and $A$ Schwinger terms does not work,
generally speaking, in the scalar case.

The first attempt in this
direction was the proposal to replace the current algebra by the
algebra of gauge fields~\cite{wein4}, the Schwinger terms in the
latter are explicitly known $c$-numbers. Considerable progress in
understanding the Weinberg sum rules was achieved due to the
introduction of Wilson's Operator Product Expansion (OPE) at short
distances~\cite{wilson} as a substitute of Lagrangian models, the
new techniques was applied to the sum rules
already in the original paper~\cite{wilson}.
The Wilson's proposal happened to be very useful tool for analysis of
convergence of spectral-function sum rules. Wilson proved, for
instance, that if his OPE method is true, the necessary and
sufficient condition for the validity of Weinberg sum rules is
that the chiral $SU(2)\times SU(2)$ group be an exact invariance
(see also~\cite{beg} for a more general proof). The discovery of
asymptotic freedom in non-Abelian gauge theories~\cite{asfr}
inspired to consider the status of spectral-function sum rules
within the framework of asymptotically free
theories~\cite{borchardt}, in particular, exploiting the
techniques
of Wilson's OPE~\cite{hagiwara}. A general recipe for
extracting exact spectral-function sum rules for asymptotically
free theories was presented in~\cite{wein5}.

Weisberger showed~\cite{weisberger} that the Wilson's results can
be reproduced in Lagrangian theory by means of the use of the
renormalization-group equation. In particular, the general
theoretical criterium for the validity of spectral sum rules was
derived: the symmetry-breaking terms in a renormalizable
Lagrangian must have canonical dimensions $\delta\leq3$. Such a
soft symmetry breaking can be implemented in practice by mass
terms and by scalar fields with nonvanishing vacuum expectation
value. Provided this condition is satisfied, the propagators
approach their symmetric values in the asymptotic spacelike limit.
Choosing then certain linear combinations of propagators whose
asymptotics will be less singular than that of the individual terms
and applying the spectral representations for those combinations,
one obtains precisely the Weinberg sum rules. To realize the
program completely, one needs to know nonleading asymptotic terms
which depend on the symmetry breaking effects. The earlier studies
of chiral symmetry breaking within various spectral-function sum
rules can be found in~\cite{das2,borchardt,hagiwara,prasad}.

In summary, it became clear that the two Weinberg sum rules are
very general and do not depend on the dynamics of chiral symmetry
breaking in the vacuum for the asymptotically free theories, such
as QCD, while the higher-order sum rules do depend on that
dynamics, hence, on the specific details of the QCD Lagrangian.

The invention of famous ITEP sum rules~\cite{svz} was
a significant progress in development of OPE-based sum rules.
Within that method, one improves the convergence and suppresses the
contribution of higher excitations by performing the Borel transformation
in Euclidean space, and then one parametrizes all non-perturbative effects
by means of a few condensates entering the numerators of OPE. As a
result, the perturbative and non-perturbative contributions become
effectively factorized, which permits to make numerous predictions
having at hand only several inputs --- the phenomenological values
of condensates. There is still no complete understanding why this
method works so well in the phenomenology. In particular, the
application of the same method to a solvable quantum-mechanical
problem~\cite{lucha} shows that when the hadron continuum is not
known and is modelled by an effective continuum threshold (this
very situation one has usually in practice), the systematic
uncertainties of the method cannot be controlled. The assumption
of dominance of the lowest-lying resonances is sometimes also in
doubt~\cite{moussallam}. A possibility to take into account the
higher excitations appeared within the Finite Energy Sum Rules
(FESR)~\cite{fesr} (see, e.g.,~\cite{fesr1} for references),
which represent a kind of extension of the Weinberg sum
rules based on analytic properties of correlation functions and on
quark-hadron duality.

About twenty years ago it was realized that the sum rules can be
directly confronted with experiment through semi-leptonic
$\tau$-lepton decays, namely the $V$ and $A$ spectral
functions can be reconstructed in the kinematical range limited by
the $\tau$-lepton mass. The first attempt was undertaken
in~\cite{argus1} using the ARGUS data~\cite{argus}. This analysis
was followed by improved versions~\cite{argus2}. Subsequently,
much more precise data of the ALEPH and OPAL collaborations on the
$V$ and $A$ spectral functions~\cite{alephopal} gave rise
to a large series of papers devoted to the extraction of
hadronic parameters, such as condensates, with the help of FESR
and other methods~\cite{aleph1,aleph2,aleph3,aleph4,aleph5}.

In the last decade the sum rules saturated by narrow resonances have
found numerous applications in the phenomenology, it would require a
long paper to survey this activity. In this respect, a question
might even appear whether new papers on such well known sum rules
are really needed. We believe, however, that some new trends in
the phenomenology invite to return to foundations of resonance sum
rules and to revise them. Each scheme of resonance saturation
implies a certain pattern for the chiral symmetry breaking at low
energies, the most known pattern is based on the conception that
the $a_1$-meson is the chiral partner of the $\rho$-meson and the
$\sigma$-meson is that of pion. If the Wigner-Weyl realization of
chiral symmetry was somehow restored maintaining confinement in
QCD, these chiral partners would be degenerate. The same pattern
is used for construction of many effective quark models, equal
resonance content provides a possibility to match these models to
QCD sum rules establishing thereby a correspondence of effective
models to the fundamental theory (see, e.g.,~\cite{effmod,AnE} and references
therein). There is, however, an alternative possibility~\cite{w166}
where the $\rho$-meson is a "would-be" chiral partner of pion. It
is not excluded that this pattern would be preserved even if the
chiral symmetry was restored --- the so-called vector
manifestation scenario~\cite{HY}. To a certain extent, the recent
phenomenological observations yield an unexpected support for this
scenario --- the $\rho$-meson belongs to the leading Regge
trajectory, the states lying on such trajectories, probably, do not
have parity partners~\cite{plb06,mpla,sv} and if the chiral symmetry is
effectively restored above the chiral symmetry breaking scale, the
chiral partner of the $a_1$-meson seems to be the
$\rho(1450)$-resonance, the first "radial" excitation of the
$\rho$-meson (such a possibility was explored in~\cite{AnE}). This
example shows that further systematic studies of both saturation
schemes and relations between the QCD sum rules and
phenomenological Lagrangians are needed, we will address to these
subjects.

Recently the resonance sum rules were employed to demonstrate that
the chiral symmetry is realized in the Wigner-Weil mode in the
upper part of meson spectrum~\cite{beane,we,sh}, the first attempt of this kind
seems to go back to~\cite{chiu}, where the baryon sector was
analyzed. These attempts boiled down to justification of parity
doubling among the highly excited states, the procedure turned out
to be model-dependent, thus not replying unambiguously whether the
chiral symmetry is restored or not. Another
approach was put forward in~\cite{npb}, where the spectrum was
split into the "chirally symmetric" and "nonsymmetric" parts. The
first part, after summation over resonances and comparison with
the OPE, yields no contribution to the condensates responsible for
the chiral symmetry breaking. If the chiral symmetry gets restored,
the second part has to represent the asymptotically vanishing
corrections to the first part. Technically, one should fix an
ansatz for mass spectrum, e.g. take the linear one, calculate its
input parameters from the imposed constraints, and compare with
the experimental data and known theoretical relations to check
whether this works. Unfortunately, the existing uncertainties both
in the experimental data and in the OPE condensates do not permit
to perform this reliably.

The chiral symmetry restoration still remains a rather iffy
concept~\cite{sv,jaffe} in spite of all efforts to justify it~\cite{glozrev},
for this reason we would prefer to use the term "parity doubling", the
latter is easier to compare with the actual spectroscopy. As
parity doubling seems to be an important observable
phenomenon~\cite{pd}, it is interesting to consider to what extent
the approximate parity degeneracy can take place in the sum rules
saturated by finite number of resonances (leaving aside the case
of trivial degeneracy) and check, if possible, whether such
mass spectra are more preferable phenomenologically in comparison
with mass spectra without approximate parity doubling. This
subject will be also addressed in the present work.

The paper is organized as follows. In Sect.~2 we concern a
relation between the generalized Weinberg sum rules and the
phenomenological Lagrangians of effective field theory. Sect.~3
deals with the same sum rules in the correlator approach. Sect.~4
is devoted to solutions of sum rule equations at different
saturation schemes and additional assumptions, with the main
emphasis being placed on the possibility for parity doubling. In
Sect.~5 we comment on some problems emerging in the sum rules with
infinite number of states. Our conclusions are summarized in
Sect.~6.

\section{Sum Rules: Lagrangian Approach}

Deriving the higher-order Weinberg sum rules
from the OPE for correlation functions, one typically encounters
the following problem: From the OPE side, the condensate terms
have anomalous dimensions, while from the resonance side, the
anomalous dimensions are absent by construction, thus the question
arises about the correctness of equating both sides. In this
section, we will argue that the sum rules are closely related to
the Lagrangian approach, in fact they can be derived from this
approach assuming the asymptotic chiral symmetry restoration at
large momentum transfer. The problem with anomalous dimensions can
be escaped in this case.

At present the approximation of narrow resonances has a solid
theoretical foundation --- it is equivalent to the large-$N_c$
(or planar) limit of QCD~\cite{hoof}, where the quantity
$g^2N_c$ is kept fixed, $g$ is the QCD coupling constant.
In the planar limit, the meson states are narrow (the meson
decay width behaves as $\Gamma\sim O(1/N_c)$) and weakly
interacting.
Confining ourselves to this limit, we may therefore
regard the mesons as almost free particles.
In addition, we may suppose that all states with fixed quantum
numbers are generated by a universal external source.
The Lagrangian of free vector and axial-vector fields
generated by external sources $J^{V,A}$ is,
\be
\label{La}
L=\sum_{n}\sum_{\varphi=V,A}\left\{\frac14\left(\partial_{\mu}\varphi_{n,\nu}^a-\partial_{\nu}\varphi_{n,\mu}^a\right)^2
-\frac12m_{\varphi,n}^2\left(\varphi_{n,\mu}^a\right)^2+\varphi_{n,\mu}^aJ_{\mu}^{\varphi,a}\right\}.
\ee
Here $a$ refers to the isospin index. For the time being we neglect the Chiral Symmetry Breaking (CSB). Assume that each
field in~\eqref{La} corresponds to a conserved current. There exists then identity "current = field" in the sense that  they
act identically in matrix elements. Define the corresponding currents as
\be
\label{concur}
j_{n,\mu}^{\varphi,a}=F_{\varphi,n}m_{\varphi,n}\varphi_{n,\mu}^a,
\ee
where $F_{\varphi,n}$ are the electromagnetic decay constants. The conservation of
current~\eqref{concur} can be easily checked by taking derivative
from the corresponding equation of motion for
Lagrangian~\eqref{La} using $\partial_{\mu}J_{\mu}^{\varphi,a}=0$
due to the isospin conservation. This is equivalent to the Lorentz
gauge condition, $\partial_{\mu}\varphi_{n,\mu}^a=0$, which is
needed to obtain the standard Klein-Gordon-Fock equation with
external source,
\be
\left(\square-m_{\varphi,n}^2\right)\varphi_{n,\mu}^a=-J_{\mu}^{\varphi,a}.
\ee
Now we can construct a conserved current to which all states with fixed quantum
numbers are coupled,
\be
\label{current}
j_{\mu}^{\varphi,a}=\sum_n j_{n,\mu}^{\varphi,a}.
\ee
The fields $\varphi_{n,\mu}^a$ correspond to physical states $\Phi_n^a$, which are created when one acts by
current~\eqref{current} on the physical vacuum.
Since the latter is electrically neutral, the procedure has to be realized for
the third isospin component, the normalized expression is
\be
\label{me}
\langle0|j_{\mu}^{\varphi,3}(x)|\Phi_n^3\rangle=F_{\varphi,n}m_{\varphi,n}\frac{e^{ipx}}{\sqrt{2p_0}}\,\epsilon_{\mu}.
\ee
The decay constant $F_{\varphi,n}$ may be related to observable quantities. For instance, if we associate $V_1$
with the $\rho$-meson, matrix element~\eqref{me} can be estimated from the decay
$\rho^0\rightarrow e^+e^-$ (see Eq.~\eqref{two}).

In real world the CSB occurs, this phenomenon was proven rigorously in the large-$N_c$ limit of QCD under
some assumptions~\cite{coleman}. The main impact of CSB for us is the appearance of pions, as a result the axial
current is not conserved anymore. Let us introduce the pions by means of the Partial Conservation of Axial
Current (PCAC) hypothesis, i.e. we perform the shift
\be
A_{1,\mu}^a\longrightarrow A_{1,\mu}^a-f_{\pi}\partial_{\mu}\pi^a,
\ee
in the expressions above. Here $f_{\pi}$ is the weak pion decay
constant, $f_{\pi}=92.4$~MeV, in the chiral limit $f_{\pi}\approx87$~MeV.
The corresponding axial current is then conserved in the chiral limit, $m_{\pi}=0$,
which we shall adopt. Furthermore, we will make use of quite common assumption of generalized PCAC, i.e. the
covariant derivative of pion field is mixed with the axial field $A_{1,\mu}^a$ only, the axial fields $A_{n,\mu}^a$, $n>1$,
are supposed to correspond to heavier states ("radial" excitations).

The part of Lagrangian~\eqref{La} related to the interaction with the external sources can be rewritten (factor
constant) as
\be
\label{lag}
L_{\text{int}}=J_{\mu}^{V,a}j_{\mu}^{V,a}+J_{\mu}^{A,a}j_{\mu}^{A,a}.
\ee
Since only planar diagrams survive in the large-$N_c$ approximation, to the leading order in the large-$N_c$ counting
the amplitudes calculated from~\eqref{lag} will be saturated by the one-particle exchanges,
the relevant diagrams are displayed in Fig.~1.
\begin{figure}
\vspace{-5 cm}
\hspace{-3.9 cm}
\includegraphics[scale=1]{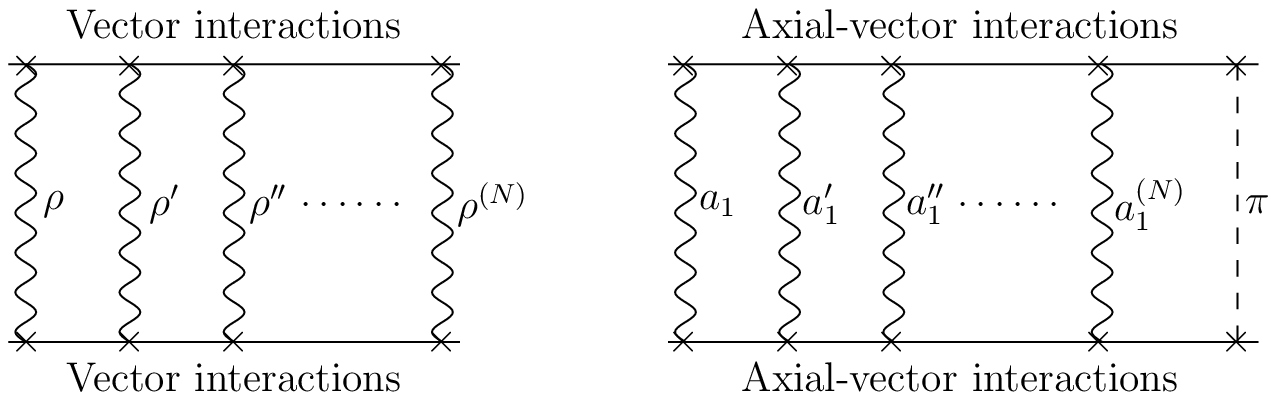}
\vspace{-21.7 cm}
\caption{\label{f1} The diagrams of meson exchanges for Lagrangian~\eqref{lag}, where the vector and axial-vector
states are supposed to be the $\rho$ and $a_1$ mesons, respectively. The "radial" excitations are marked
by primes.}
\end{figure}

The  Fourier transforms of the corresponding vector and axial-vector amplitudes are
\be
W_V(p)=CJ_{\mu}^{V,a}(p)\left(\sum_nF_{V,n}^2m_{V,n}^2\frac{-g_{\mu\nu}+\frac{p_{\mu}p_{\nu}}{m_{V,n}^2}}{p^2-m_{V,n}^2+
i\varepsilon}\right)J_{\nu}^{V,a}(p),
\ee
\be
W_A(p)=CJ_{\mu}^{A,a}(p)\left(\sum_nF_{A,n}^2m_{A,n}^2\frac{-g_{\mu\nu}+\frac{p_{\mu}p_{\nu}}{m_{A,n}^2}}{p^2-m_{A,n}^2+
i\varepsilon}+\frac{f_{\pi}^2p_{\mu}p_{\nu}}{p^2+i\varepsilon}\right)J_{\nu}^{A,a}(p),
\ee
where $C$ is a constant. Let us separate the transverse and longitudinal parts with the help of identity
\be
\frac{-g_{\mu\nu}+\frac{p_{\mu}p_{\nu}}{m^2}}{p^2-m^2+i\varepsilon}=
\frac{-g_{\mu\nu}+\frac{p_{\mu}p_{\nu}}{p^2}}{p^2-m^2+i\varepsilon}+\frac{p_{\mu}p_{\nu}}{m^2p^2}+\mathcal{O}(\varepsilon),
\ee
which leads to
\begin{multline}
W_V(p)=CJ_{\mu}^{V,a}(p)\left\{\sum_nF_{V,n}^2m_{V,n}^2\frac{-g_{\mu\nu}+\frac{p_{\mu}p_{\nu}}{p^2}}{p^2-m_{V,n}^2+
i\varepsilon}\right.+\\
+\frac{p_{\mu}p_{\nu}}{p^2}\left.\sum_nF_{V,n}^2\right\}J_{\nu}^{V,a}(p)+\mathcal{O}(\varepsilon),
\label{ampV}
\end{multline}
\begin{multline}
W_A(p)=CJ_{\mu}^{A,a}(p)\left\{\sum_nF_{A,n}^2m_{A,n}^2\frac{-g_{\mu\nu}+\frac{p_{\mu}p_{\nu}}{p^2}}{p^2-m_{A,n}^2+
i\varepsilon}\right.+\\
+\left.\frac{p_{\mu}p_{\nu}}{p^2}\left(\sum_nF_{A,n}^2+f_{\pi}^2\right)\right\}J_{\nu}^{A,a}(p)
+\mathcal{O}(\varepsilon).
\label{ampA}
\end{multline}

The exact chiral symmetry implies that the expressions in the
braces of Eqs.~\eqref{ampV} and~\eqref{ampA} have to be equal at
all $p$. This is possible only if $m_{V,n}=m_{A,n}$,
$F_{V,n}=F_{A,n}$, $f_{\pi}=0$, which is far from reality.
Instead of exact chiral symmetry, one usually imposes an
asymptotic chiral symmetry at large energies.

Let us interpolate the difference of vector and axial-vector
amplitudes at large four-momentum $|p|$ by the following Taylor
expansion,
\be
\label{diff}
W_V(p)-W_A(p)=\sum_{k=0}^{\infty}\frac{\Delta_k}{p^{2k}}.
\ee
We have taken into account that the Lorentz invariance dictates the
dependence on $p^2$ in the final answer. The quantities $\Delta_k$
are unknown constants to be determined. The requirement that the
vector and axial-vector interactions are indistinguishable at
large $p$ leads to $\Delta_0=0$, hence, to the equality of
longitudinal parts in Eqs.~\eqref{ampV} and~\eqref{ampA},
this is the first nontrivial constraint from the asymptotic chiral
symmetry. Consider difference~\eqref{diff} at
$\mathcal{O}(p^{-2})$. In the limit of very large $p$, this
difference is identical to the difference of vector and axial-vector
amplitudes in a theory where only massless vector and axial-vector
mesons are exchanged. If the vector and axial-vector interactions
are indistinguishable at large $p$ in the latter theory, the
difference above is zero. This analogy suggests that we may assume
a more strong asymptotic chiral symmetry and admit $\Delta_1=0$.
Comparing Eqs.~\eqref{ampV} and~\eqref{ampA} with
Eq.~\eqref{diff}, the following set of asymptotic sum rules can be
written,
\begin{eqnarray}
\label{w1}
\sum_nF_{V,n}^2-\sum_nF_{A,n}^2-f_{\pi}^2&=&0,\\
\label{w2}
\sum_nF_{V,n}^2m_{V,n}^2-\sum_nF_{A,n}^2m_{A,n}^2&=&0,\\
\label{w3}
\sum_nF_{V,n}^2m_{V,n}^{2k}-\sum_nF_{A,n}^2m_{A,n}^{2k}&=&\Delta_k,\qquad
k=2,3,4,\dots
\end{eqnarray}
Sum rules~\eqref{w1} and~\eqref{w2} are the Weinberg sum rules~\cite{wein}
generalized to the case of arbitrary number of states. Some
general properties of set of equations~\eqref{w1}-\eqref{w3} were
studied in~\cite{knecht}, where these sum rules were derived from
the OPE of two-point quark current
correlators~\cite{svz}. It should be emphasized the distinction of
the OPE-based ideology from the one adopted here. The quantities
$\Delta_k$ are usual finite numbers in our approach, while in the
OPE they are related to the condensates of appropriate dimension
multiplied by the coefficient functions calculated from QCD by
means of the perturbation theory\footnote{The ALEPH/OPAL data on
the $V$ and $A$ spectral functions from $\tau$ decays was used for
numerical estimations of quantities $\Delta_k$ up to dimension 18,
see~\cite{aleph4} for a review.}. The relation with the
fundamental theory is an advantage of the OPE-based methods, in
particular, the restrictions $\Delta_0=0$ and $\Delta_1=0$ follow
automatically in the chiral limit. However, from the point of view
of sum rules~\eqref{w3}, the OPE has two shortcomings. First, the
OPE represents, at best, an asymptotic expansion with zero radius of
convergence, therefore the calculation of $\Delta_k$ at large $k$
(in practice, at $k\gtrsim4$) is not reliable because the
divergence sets in. Second, as mentioned above the condensate terms
have an anomalous dimension while the l.h.s. of Eq.~\eqref{w3}
does not have, this circumstance caused a critics of such sum
rules recently~\cite{peris}.

Another derivation of generalized Weinberg sum rules, Eqs.~\eqref{w1}
and~\eqref{w2}, which does not use the correlation functions, was
proposed in~\cite{beane}, where the method consisted in analysis
of certain Current Algebra commutation relations in infinite momentum
frame.

\section{Sum Rules: Correlator Approach}

Originally the sum rules under consideration were derived in~\cite{wein}
from some asymptotic restrictions on correlation functions with
subsequent saturation by narrow states\footnote{An earlier
application of asymptotic equality for the $V$ and $A$ correlators
exists in the literature~\cite{gasiorowicz}, where the rate for
$\omega^0\rightarrow\pi^0+\gamma$ was calculated from one of sum
rules emerging in the infinite energy limit. We are grateful to
Prof.~S.~Gasiorowicz for this remark.}. We will
consider a generalization of this method to the case of arbitrary
number of narrow states and discuss the underlying physics.

Introduce the vector and axial-vector two-point correlation functions,
\be
\label{veccor}
\Pi_{\mu\nu}^V(p)=i\int
d^4x\,e^{ipx}\left\langle0\left|T\left(j_{\mu}^{V,a}(x)j_{\nu}^{V,a}(0)\right)\right|0\right\rangle,
\ee
\be
\Pi_{\mu\nu}^A(p)=i\int
d^4x\,e^{ipx}\left\langle0\left|T\left(j_{\mu}^{A,a}(x)j_{\nu}^{A,a}(0)\right)\right|0\right\rangle.
\ee
The $V$ and $A$ currents are commonly interpolated by the
following quark bilinears,
\be
\label{currents}
j_{\mu}^{V,a}=\bar{q}\gamma_{\mu}\frac{\tau^a}{2}q,\qquad
j_{\mu}^{A,a}=\bar{q}\gamma_{\mu}\gamma_5\frac{\tau^a}{2}q,
\ee
where $\tau^a$ are isospin Pauli matrices. For the time being we
do not need explicit expression for these currents.
Consider the spectral representation for the vector correlator,
\be
\label{vecsp}
\Pi_{\mu\nu}^V(p)=\int\limits_{0}^{\infty} dm^2\,\rho^V(m^2)\frac{-g_{\mu\nu}+\frac{p_{\mu}p_{\nu}}{m^2}}{p^2-m^2+
i\varepsilon}+\text{S.T.},
\ee
where $\rho^V(m^2)$ is spectral density, meaning an amplitude of probability that external vector field $V$ creates
an excitation with the mass $m$. The letters S.T. denote a Schwinger term, which emerges because the spectral
representation is covariant, while the chronological $T$-product in Eq.~\eqref{veccor} is not covariant~\cite{sak,adl}.
Separating the transverse and longitudinal parts and substituting a manifest expression for the Schwinger term
in our case~\cite{sak}, we obtain,
\begin{multline}
\label{15}
\Pi_{\mu\nu}^V(p)=\int\limits_{0}^{\infty} dm^2\,\rho^V(m^2)\frac{-g_{\mu\nu}+\frac{p_{\mu}p_{\nu}}{p^2}}{p^2-m^2+
i\varepsilon}
+\frac{p_{\mu}p_{\nu}}{p^2}\int\limits_{0}^{\infty} dm^2\,\frac{\rho^V(m^2)}{m^2}+\\
+\delta_{0\mu}\delta_{0\nu}\int\limits_{0}^{\infty} dm^2\,\frac{\rho^V(m^2)}{m^2}+\mathcal{O}(\varepsilon).
\end{multline}
Analogous expression takes place for the axial-vector case,
\begin{multline}
\label{16}
\Pi_{\mu\nu}^A(p)=\int\limits_{0}^{\infty} dm^2\,\rho^A(m^2)\frac{-g_{\mu\nu}+\frac{p_{\mu}p_{\nu}}{p^2}}{p^2-m^2+
i\varepsilon}
+\frac{p_{\mu}p_{\nu}}{p^2}\left(\int\limits_{0}^{\infty} dm^2\,\frac{\rho^A(m^2)}{m^2}+f_{\pi}^2\right)+\\
+\delta_{0\mu}\delta_{0\nu}\left(\int\limits_{0}^{\infty} dm^2\,\frac{\rho^A(m^2)}{m^2}+f_{\pi}^2\right)+
\mathcal{O}(\varepsilon).
\end{multline}

Let us require the asymptotic chiral symmetry~\cite{das2},
\be
\label{asymsym}
\lim_{|p|\rightarrow\infty}\left(\Pi_{\mu\nu}^V(p)-\Pi_{\mu\nu}^A(p)\right)=0.
\ee
The first Weinberg sum rule then follows from the longitudinal part $p_{\mu}p_{\nu}/p^2$,
\be
\int\limits_{0}^{\infty} dm^2\,\frac{\rho^V(m^2)}{m^2}=\int\limits_{0}^{\infty} dm^2\,\frac{\rho^A(m^2)}{m^2}+f_{\pi}^2.
\ee
The assumption that chiral symmetry is stronger leads to nullifying the transverse part at
$(-g_{\mu\nu}+p_{\mu}p_{\nu}/p^2)/(p^2-m^2)$, this is the second Weinberg sum rule,
\be
\label{sws}
\int\limits_{0}^{\infty} dm^2\,\rho^V(m^2)=\int\limits_{0}^{\infty} dm^2\,\rho^A(m^2).
\ee

In general case, the transverse parts satisfy the following condition,
\be
\lim_{p\rightarrow\infty}p^{2k}\left(\Pi_{\mu\nu}^V(p)-\Pi_{\mu\nu}^A(p)\right)=\Delta_k,
\qquad k=1,2,\dots,
\ee
where $\Delta_k$ are some constants of mass dimension $2k+2$. The corresponding
asymptotic sum rules are
\be
\label{gensr}
\int\limits_{0}^{\infty} dm^2\,m^{2(k-1)}\rho^V(m^2)-\int\limits_{0}^{\infty} dm^2\,m^{2(k-1)}\rho^A(m^2)=\Delta_k,
\ee
which reduce to the second Weinberg sum rule~\eqref{sws} for $k=1$, $\Delta_1=0$.

Using the definition of spectral density,
\be
\label{spd}
\left(-g_{\mu\nu}+\frac{p_{\mu}p_{\nu}}{p^2}\right)\rho(p^2)=
(2\pi^3)\sum_n\delta^{(4)}(p-p_n)\langle 0|j_{\mu}^3(0)|n\rangle\langle n|j_{\nu}^3(0)|0\rangle,
\ee
one can saturate the obtained equations by narrow resonances,
\be
\rho^{V,A}(m^2)=\sum_n
F_{V,A;n}^2m^2_{V,A;n}\delta\left(m^2-m^2_{V,A;n}\right),
\ee
and get the sum rules considered above, Eqs.~\eqref{w1}-\eqref{w3}.

An interesting consequence of the correlator formalism is that the first
Weinberg sum rule follows also from the requirement of cancellation of
Schwinger terms in the $\Pi_V-\Pi_A$ difference~\cite{sak}, this fact is evident
from Eqs.~\eqref{15} and~\eqref{16}. The cancellation of Schwinger terms
is quite natural as long as the $\Pi_V-\Pi_A$ difference is related to observable
quantities. Thus, one arrives at a kind of equivalence between the
asymptotic chiral symmetry and the cancellation of Schwinger terms. A question
emerges, which principle is more fundamental?

The Schwinger terms emerge in equal-time current commutators when one defines a current as the limit
\be
j_{\Gamma}(x_0,{\bf x})\equiv\lim_{\varepsilon\rightarrow0}\bar{q}(x_0,{\bf x+\varepsilon})\Gamma
q(x_0,{\bf x-\varepsilon}),
\ee
where $j_{\Gamma}$ is the current corresponding to the gamma-matrix structure $\Gamma$. These terms
can be both operators and $c$-numbers, generally speaking, their form is model-dependent. The appearance
of Schwinger terms is an inescapable consequence of Lorentz invariance and of positive definiteness
for probability, otherwise the theory is trivial~\cite{sak,adl}.
This observation suggests that the cancellation of Schwinger terms is likely
more fundamental requirement than the cancellation of the longitudinal parts
at large momentum. In addition, the former cancellation should take place at all
momenta as it is momentum-independent. The given property explains why
the first Weinberg sum rule is much better fulfilled in the phenomenology
than the second one, the reason seems to be that it is not only asymptotic
sum rule --- its validity extends beyond the high-energy domain.
Indeed, provided the equality of Schwinger terms in spectral
representation~\eqref{vecsp} and in its axial-vector analogue, the first
Weinberg sum rule equally emerges if one takes the limit $|p|\rightarrow0$ in
condition~\eqref{asymsym}.

Asymptotic sum rules at large momentum can be supplemented by low-energy
sum rules at vanishing momentum, which are nothing but an extension
of sum rules~\eqref{gensr} to negative $k$ provided by the requirement
\be
\lim_{|p|\rightarrow0}p^{2k}\left(\Pi_{\mu\nu}^V(p)-\Pi_{\mu\nu}^A(p)\right)=\Delta_k,
\qquad k=-1,-2,\dots
\ee
The constant $\Delta_{-1}$ is known from the phenomenology~\cite{ecker},
\be
\Delta_{-1}=-4\bar{L}_{10},
\ee
where $\bar{L}_{10}$ is the scale independent part of the coupling
of the relevant operator in the $\mathcal{O}(p^4)$ effective chiral
Lagrangian of QCD~\cite{gasser}. This constant can be expressed
by means of the following combination of hadronic parameters,
\be
\label{Dsr}
\bar{L}_{10}=-\frac14\left(\frac13 f_{\pi}^2
\langle r_{\pi}^2\rangle-\mathcal{F}_A\right),
\ee
where $\langle r_{\pi}^2\rangle$ is the electromagnetic mean mass
squared radius of the charged pions and $\mathcal{F}_A$ is the axial-vector
coupling. Relation~\eqref{Dsr} is nothing but the Das-Mathur-Okubo
low-energy theorem~\cite{das3}.

A general property of all considered sum rules is that they depend on the
four-momentum cutoff $\mu$ through the number of included resonances
only. Due to the equality $\Delta_1=0$ in the second Weinberg sum rule,
another sum rule with the given property can be written,
\be
\int\limits_{0}^{\infty}dm^2\,m^2
\ln\left(\frac{m^2}{\mu^2}\right)\rho^V(m^2)-
\int\limits_{0}^{\infty}dm^2\,m^2
\ln\left(\frac{m^2}{\mu^2}\right)\rho^A(m^2)=\bar{\Delta}_1.
\ee
The constant $\bar{\Delta}_1$ turns out to be related with a remarkable
physical observable, the electromagnetic pion mass difference~\cite{das},
$\bar{\Delta}_1\sim m_{\pi^{\pm}}^2-m_{\pi^0}^2$, we will exploit this
relation later.

\section{Weinberg-like Sum Rule Equations: Solutions for Particular Cases}

We are going to solve the system of
equations~\eqref{w1}-\eqref{w3} for some typical cases. Partly,
our analysis may be regarded as a revision and extension of
results obtained in~\cite{knecht}. We, however, place quite
different emphasis --- we are interested in maximally degenerate
solutions (aside from the trivial ones) for
excited $V$ and $A$ mesons and in overall correspondence of
different possibilities to the actual experimental data.

\subsection{One Vector and One Axial-vector}

In this subsection we revisit the standard Weinberg ansatz and
clarify its modern status.

It is convenient to work with the dimensionless quantities,
\begin{gather}
X_{V,A}=\frac{F_{V,A}^2}{f_{\pi}^2},\\
Y_{V,A}=\frac{M_{V,A}^2}{M_{\rho}^2}.
\end{gather}
Experimentally~\cite{pdg} (in MeV),
\be
\label{ed}
F_{V}=154\pm8,\quad F_{A}=123\pm25,\quad M_{V}=776,\quad M_{A}=1230\pm40,
\ee
and $f_{\pi}\approx87$~MeV in the chiral limit which will be used
in the sequel for our estimations, we remind that in real world $f_{\pi}=92.4$~MeV.
We will associate the $V$ state with the $\rho$-meson, the first three sum rules
look then as follows,
\be
\label{35}
\left\{
\begin{aligned}
X_{\rho}-X_A&=1\\
X_{\rho}-X_AY_A&=0\\
X_{\rho}-X_AY_A^2&=\Delta,\\
\end{aligned}
\right.
\ee
where the notation $\Delta=\Delta_2$ is adopted (see
Eq.~\eqref{w3}).
The first two sum rules were first proposed by
Weinberg~\cite{wein}. They give for the third sum rule,
\be
\label{delt}
\Delta=X_{\rho}(1-Y_A)=-\frac{X_{\rho}}{X_{\rho}-1}.
\ee
From OPE~\cite{svz},
\be
\Delta_{\text{OPE}}=-\frac{4\pi\alpha_s\langle\bar{q}q\rangle^2}{M_{\rho}^4f_{\pi}^2}.
\ee
At 1 GeV one has for the strong coupling and the quark condensate,
\be
\alpha_s\approx0.5,\qquad \langle\bar{q}q\rangle\approx
(-235\,\text{MeV})^3.
\ee
The phenomenological and experimental values above yield the
following estimate,
\be
\Delta_{\text{OPE}}\approx-0.3.
\ee

Another two sum rules follow from the electromagnetic pion mass difference~\cite{das},
\be
\label{emdif}
m_{\pi^{\pm}}-m_{\pi^0}=-\frac{C_0}{f_{\pi}^2}\int\limits_{0}^{\infty}dm^2\,m^2
\ln\left(\frac{m^2}{\mu^2}\right)\left(\rho^V(m^2)-\rho^A(m^2)\right),
\ee
where
\be
C_0=\frac{3\alpha}{8\pi m_{\pi}},
\ee
($\alpha$ is the fine structure constant and $\mu$ denotes an arbitrary scale)
and the scale independent constant $\bar{L}_{10}$ of effective Chiral Lagrangian~\cite{gasser},
\be
\label{Lten}
\bar{L}_{10}=-\frac14\int\limits_{0}^{\infty} dm^2\,\frac{\rho^V(m^2)-\rho^A(m^2)}{m^2}.
\ee
The experiment~\cite{pdg} and the chiral phenomenology~\cite{pich} yield
for $m_{\pi^{\pm}}-m_{\pi^0}$ and $\bar{L}_{10}$ respectively
\be
m_{\pi^{\pm}}-m_{\pi^0}=4.6\,\text{MeV},\qquad
\bar{L}_{10}=-(5.5\pm0.7)\cdot10^{-3}.
\ee

Saturating the spectral densities by one resonance plus continuum
in~Eqs.~\eqref{emdif} and~\eqref{Lten}, we obtain
\begin{eqnarray}
m_{\pi^{\pm}}-m_{\pi^0}&=&-\frac{C_0}{f_{\pi}^2}\left(M_{\rho}^2F_{\rho}^2\ln\frac{M_{\rho}^2}{\mu^2}-
M_A^2F_A^2\ln\frac{M_A^2}{\mu^2}\right)\nonumber\\
&=& C_0M_{\rho}^2X_{\rho}\ln\frac{X_{\rho}}{X_{\rho}-1},
\label{44}
\end{eqnarray}
\be
\label{45}
\bar{L}_{10}=-\frac{f_{\pi}^2}{4M_{\rho}^2}\left(\frac{X_{\rho}}{Y_{\rho}}-\frac{X_A}{Y_A}\right)=
-\frac{f_{\pi}^2}{4M_{\rho}^2}\left(2-\frac{1}{X_{\rho}}\right),
\ee
where the first and the second Weinberg sum rules have been used.

To predict concrete numerical values one should fix an input
parameter, say $X_{\rho}$. Originally~\cite{wein} Weinberg assumed the KSFR
relation~\cite{ksfr},
\be
X_{\rho}=2,
\ee
resulting in
\be
X_{A}=1,\qquad Y_A=2.
\ee
We will refer to this ansatz as the "Weinberg" one. It predicts (in the
chiral limit),
\be
\Delta=-2,\qquad \bar{L}_{10}\approx-4.7\cdot10^{-3},
\ee
and (in MeV),
\be
F_{\rho}\approx123,\quad F_A\approx87,\quad M_A\approx1100,\quad
m_{\pi^{\pm}}-m_{\pi^0}\approx5.2.
\ee
Identifying the $A$ state with the $a_1$-meson,
the Weinberg ansatz was widely used in the literature for matching
conditions and other purposes.

We would like to note, however, that the status of the KSFR
relation caused much discussions in the literature. In particular,
its original derivation was revised in~\cite{ksfr2}, the main
lesson was that it required more {\it ad hoc} assumptions in
comparison with the ones made in~\cite{ksfr}.
Notably, the modern experimental data~\eqref{ed}
favors rather to the ansatz
\be
X_{\rho}=3,
\ee
leading to
\be
X_{A}=2,\qquad Y_A=\frac32.
\ee
We will refer to this ansatz as the "experimental" one. It predicts (in the
chiral limit),
\be
\Delta=-\frac32,\qquad L_{10}\approx-5.2\cdot10^{-3},
\ee
and (in MeV),
\be
F_{\rho}\approx151,\quad F_A\approx123,\quad M_A\approx950,\quad
m_{\pi^{\pm}}-m_{\pi^0}\approx4.6.
\ee
Comparing the predictions with the corresponding experimental and
phenomenological values given above, one can see immediately that
the experimental ansatz works better substantially than the
Weinberg one: Lowering $M_A$ by 15\% (which is within the
large-$N_c$ accuracy), one amends noticeably all other five quantities.

Finally, it should be mentioned that the degenerate case, which
would mean here $Y_A=1$, cannot be obtained within the considered ansatz.

\subsection{One Vector and No Axial-vector}

As was discussed in Introduction, it makes sense to consider the
ansatz without the axial-vector meson, the "would-be" chiral
partner of the vector state is then the pion (the given possibility
was not analyzed in~\cite{knecht}). This case amounts to
setting $X_A=0$ in the formulae of the previous subsection. First
of all, it is obvious that only the first sum rule in
Eqs.~\eqref{35} can be satisfied with this ansatz, yielding
$X_{\rho}=1$, i.e. $F_{\rho}=f_{\pi}$. The physical sense of this
result is that had the "vector manifestation" been exactly
realized, only the first sum rule would have survived because the
$\rho$-meson is then massless. The value of constant
$\bar{L}_{10}$ from Eq.~\eqref{45} is
\be
\bar{L}_{10}=-\frac{f_{\pi}^2}{2M_{\rho}^2}\approx-6.3\cdot10^{-3},
\ee
which represents a quite reasonable estimate. Taking into account
$X_{\rho}=1$, now the electromagnetic mass difference from
Eq.~\eqref{44},
\be
m_{\pi^{\pm}}-m_{\pi^0}=C_0M_{\rho}^2\ln\frac{\mu^2}{M_{\rho}^2},
\ee
depends explicitly on the cutoff $\mu$. Treating the cutoff as
an input parameter, one can achieve any value for
$m_{\pi^{\pm}}-m_{\pi^0}$, in particular, the experimental one,
$4.6$~MeV, is implemented at $\bar{\mu}\approx1435$~MeV. In this respect,
it is not surprising that $m_{\pi^{\pm}}-m_{\pi^0}$ can be calculated
without use of the $a_1$-meson, see e.g.~\cite{dpihls}.
On the other hand, the
cutoff should not exceed the mass of the higher resonance,
otherwise the latter has to be included explicitly. Since the
higher resonances give a sizeable contribution to
$m_{\pi^{\pm}}-m_{\pi^0}$~\cite{dpi}, the physical interpretation
for $\bar{\mu}$ seems to be the following: It indicates indirectly
on the typical scale of the higher meson excitations such as
$a_1(1230)$ and $\rho(1450)$. The result of the previous
subsection is reproduced if we identify the cutoff with the mass
of the lowest $A$-meson.

In summary, the $\rho-\pi$ ansatz is not senseless within the sum
rules, it seems to be indeed the simplest ansatz consistent, to a
certain extent, with the phenomenology.

\subsection{Two Vectors and One Axial-vector}

The ansatz with two $V$-mesons and one $A$-state was considered
in~\cite{knecht} and also in~\cite{aleph3,AnE}. We will reexamine
this possibility following the emphasis expressed in the beginning
of this section.

Extending sum rules~\eqref{35} by the second vector state, we
arrive at
\be
\left\{
\begin{aligned}
X_{\rho}-X_A+X_V&=1\\
X_{\rho}-X_AY_A+X_VY_V&=0\\
X_{\rho}-X_AY_A^2+X_VY_V^2&=\Delta.\\
\end{aligned}
\right.
\label{syst3}
\ee
It is natural to associate the second vector state with the
$\rho(1450)$-meson, whose mass is~\cite{pdg}
\be
\label{rp}
M_V=1459\pm11\,\text{MeV},
\ee
while its decay constant $F_V$ is unknown.
The solution of system~\eqref{syst3} can be written as
\be
\left\{
\begin{aligned}
X_A&=\frac{\left((X_{\rho}-1)Y_V-X_{\rho}\right)^2}{(X_{\rho}-1)Y_V^2-2X_{\rho}Y_V+X_{\rho}-\Delta}\\
X_V&=\frac{X_{\rho}+(X_{\rho}-1)\Delta}{(X_{\rho}-1)Y_V^2-2X_{\rho}Y_V+X_{\rho}-\Delta}\\
Y_A&=\frac{X_{\rho}(Y_V-1)+\Delta}{(X_{\rho}-1)Y_V-X_{\rho}}.
\label{sol3}
\end{aligned}
\right.
\ee
These solutions are positive at
\be
\Delta\geq-\frac{X_{\rho}}{X_{\rho}-1},
\ee
with the inequality,
\be
Y_V\geq Y_A,
\ee
being always maintained.
Solutions are not valid when the corresponding denominator is
zero, the corresponding special point is (we are interested in the physical case $Y_V>1$ only),
\be
(Y_V)_s=(Y_A)_s=\frac{X_{\rho}+\sqrt{X_{\rho}+(X_{\rho}-1)\Delta}}{X_{\rho}-1}.
\ee

Physically we expect that the $\rho$-meson dominates over heavier
resonances, this leads to the inequalities
\be
X_A\leq X_{\rho}, \qquad X_V\leq X_{\rho}.
\ee
The second inequality holds automatically because
$X_V\leq1$ due to the first sum rule in system~\eqref{syst3},
while physically $X_{\rho}\geq2$. The first inequality results in
the lower bound on $Y_V$,
\be
Y_V\geq\frac{X_{\rho}}{X_{\rho}-1}\left(1+\sqrt{1+\frac{X_{\rho}-1}{X_{\rho}}\Delta}\right).
\ee
To estimate the ensuing restriction on the mass $M_V$ we can admit
$\Delta=0$. Then one has $M_V\geq1550$ MeV for $X_{\rho}=2$ and $M_V\geq1340$ MeV for
$X_{\rho}=3$. Thus, the physical value of
$\rho(1450)$-meson~\eqref{rp} mass is incompatible with the KSFR
relation within the ansatz under consideration.

The possibility $Y_V=Y_A$ is realized at finite $X_V$ and $X_A$
only if
\be
\label{62}
\Delta=-\frac{X_{\rho}}{X_{\rho}-1},
\ee
while in this case
\be
\label{63}
Y_V=Y_A=\frac{X_{\rho}}{X_{\rho}-1},
\ee
i.e. Eqs.~\eqref{62} and~\eqref{63} yield
\be
\label{rel}
\Delta+Y_A=0.
\ee

The impossibility to provide $\Delta=0$ for degenerate case in
realistic situations recurs at introducing more resonances. For
instance,
if we add a pair of resonances ($\rho''$ and $a_1'$ mesons) with
equal masses then one can show that the condition $\Delta=0$ may
be adjusted only when $X_A>2(X_{\rho}+1)$ while we expect
$X_A\leq X_{\rho}$.

Finally we note that the considered possibility for the degenerate
case, $Y_V=Y_A$, was missed in the analysis~\cite{knecht}. The
solution of system~\eqref{sol3} was written in~\cite{knecht} as
(in our notations)
\be
\left\{\begin{aligned}
X_{\rho}&=\frac{Y_AY_V+\Delta}{(Y_A-1)(Y_V-1)}\\
X_A&=-\frac{Y_V+\Delta}{(Y_A-1)(Y_V-Y_A)}\\
X_V&=\frac{Y_A+\Delta}{(Y_V-1)(Y_V-Y_A)}.
\end{aligned}
\right.
\ee
The conclusion made in~\cite{knecht} was that when
relation~\eqref{rel} takes place then $X_V=0$ and the highest vector
state decouples. Evidently, this is not true if $Y_V=Y_A$, i.e.
when the $V$ and $A$ mesons are exactly degenerate.

\subsection{Arbitrary Finite Number of States}

The case of arbitrary, but finite, number of states is very model
dependent. The general analysis of this case performed
in~\cite{knecht} was essentially based on the following
simplification: The number of unknown variables $X_{V,A}$ is equal
to the number of sum rules under consideration, the system of
equations for $X_{V,A}$ is then linear. Clearly, this is a model
assumption, the real-life physics should not depend on the way one
solves the sum rule equations. We are interested in the most
degenerate case other than the complete degeneracy of radial
excitations in masses and residues, i.e. when all is determined by
the ground states only. The physical spectrum looks like a
perturbed linear (in masses square) spectrum. Staying within the
linear parametrization, it is possible to satisfy the two Weinberg
sum rules with arbitrary number of states by fine-tuning of
residues, but hardly possible to satisfy the higher-order sum rules, this
would require the introduction of model dependent corrections to
the linear behaviour. As an example close to the real-life physics
we give the following ansatz, which satisfies the two Weinberg
sum rules identically,
\be
\label{69}
\Pi^V(p^2)=\frac{2F_{\rho}^2}{p^2-m_{\rho}^2+i\varepsilon}+
2\sum_{n=1}^{N}\frac{F_{V,n}^2}{p^2-m_{V,n}^2+i\varepsilon}+\text{P.C.},
\ee
\be
\Pi^A(p^2)=\frac{2f_{\pi}^2}{p^2+i\varepsilon}+
2\sum_{n=1}^{N}\frac{F_{A,n}^2}{p^2-m_{A,n}^2+i\varepsilon}+\text{P.C.},
\ee
where P.C. means "perturbative continuum", the $\Pi^{V,A}(p^2)$ are
defined as
\be
\label{trans}
\Pi^{V,A}_{\mu\nu}(p)=(-g_{\mu\nu}p^2+p_{\mu}p_{\nu})\Pi^{V,A}(p^2),
\ee
and the masses and residues are as follows
\be
\label{ans1}
F_{\rho}^2=2f_{\pi}^2, \qquad F_{V,n}^2=\left\{
                                        \begin{aligned}
                                         2f_{\pi}^2,&\quad n<N,\\
                                         f_{\pi}^2,&\quad n=N
                                        \end{aligned}
                                       \right.
,\qquad m_{V,n}^2=m_{\rho}^2(2+2n);
\ee
\be
\label{ans2}
F_{A,n}^2=2f_{\pi}^2, \qquad m_{A,n}^2=m_{\rho}^2(1+2n).
\ee
In this example the $\rho$-meson is singled out, its residue is in
accord with the KSFR relation, the universal slope $2m_{\rho}^2$
agrees with the phenomenology and some models (see~\cite{AE} for
discussions) as well as the universal residue $2f_{\pi}^2$. We
have made here a minimal manipulation with residues --- the
residue of the highest vector state is two times less than the
universal one. The physical interpretation could be given the
following: The resonance of mass $m_{V,N}^2=m_{\rho}^2(2+2N)$
is the heaviest in the system, if one cuts off at
$\mu_{\text{cut}}=m_{V,N}$ then the half of its decay width
(namely, the right half from the position of resonance) is
thrown away, in the narrow-width approximation this loss of
information can be mimicked by halfing the residue.

We could not achieve the completely
degenerate case, $m_{V,n}^2=m_{A,n}^2$, at least in asymptotics,
a removal of degeneracy seems to be unavoidable if one likes to get
rid of nonlinear corrections. It is interesting to note that the
mass spectrum of the ansatz above resembles that of old dual
models~\cite{avw} (a somewhat similar model, but for infinite number
of states, was considered in~\cite{plb}).

In the case of arbitrary number of states there is one subtlety
which is practically always ignored in the sum rules under
consideration --- there exist two kinds of $V$-mesons, the
$S$-wave and $D$-wave ones, and both couple to the interpolating
$V$-current~\eqref{currents}~\cite{sv}. The only exception is
paper~\cite{we}, where the problem was addressed for the case of
infinite number of resonances, but the discussions of that paper
remain relevant for our case. The doubling of $V$-states forces to
replace the sum in Eq.~\eqref{69} by the following one,
\be
\sum_{n=1}^{N}\frac{F_{V,n}^2}{p^2-m_{V,n}^2+i\varepsilon}\longrightarrow
\sum_{n=1}^{N}\frac{F_{V_S,n}^2}{p^2-m_{V_S,n}^2+i\varepsilon}+
\sum_{n=k}^{N}\frac{F_{V_D,n}^2}{p^2-m_{V_D,n}^2+i\varepsilon},
\ee
where $k$ is some integer (in practice, $k=2$ if the $\rho$-meson
is singled out, see Eq.~\eqref{nL}). As to the $D$-wave vector
states, two alternative possibilities were proposed in~\cite{we}:
(i) the $D$-wave mesons approach the $S$-wave
$V$-trajectory, implying asymptotic degeneration; (ii) $D$-wave
$V$-mesons decouple. The conclusion made in~\cite{we} was that the
possibility (ii) is more plausible. The recent phenomenological
studies, however, invite to reconsider this conclusion. Namely,
the spectrum of light nonstrange mesons  as a function of radial
$n$ and angular momentum $L$ quantum numbers seems to obey
a simple relation~\cite{sv,mpla,forkel,pd,prc},
\be
\label{nL}
m^2_{n,L}\sim n+L,
\ee
the case $L=0$ corresponds to the $S$-wave states, while for $L=2$
one has the $D$-wave mesons. It is obvious from this relation that
an approximate degeneracy of $S$- and $D$-wave states occurs,
the effect was found by the Crystal Barrel Collaboration~\cite{bugg},
although the experimental results are still preliminary.
Identifying now
\be
m_{V,n}^2=m_{V_S,n}^2, \qquad F_{V,n}^2=\left\{
                                        \begin{aligned}
                                         F_{V_S,n}^2,\phantom{{}+F_{V_D,n}^2}&\quad n<k,\\
                                         F_{V_S,n}^2+F_{V_D,n}^2,&\quad
                                         n\geq k
                                        \end{aligned}
                                       \right.,
\ee
we arrive at the standard pattern of resonance saturation, so the
usual formulae remain formally valid as if the $D$-wave states were
decoupled. A qualitative argument in favour of $D$-wave decoupling
presented in~\cite{we} relied, in essence, on the fact that the
vector interpolating current~\eqref{currents} couples to the
$e^+e^-$ annihilation, which is a point-like process, hence, the
extended objects like the $D$-wave states should decouple, i.e.
their residues vanish rapidly. However, quasiclassical arguments
(see, e.g., discussions in~\cite{sv}) tell us that the orbitally
and radially excited mesons are equally extended objects, at least
in the hadron string picture, in which the size of meson is
defined by its mass only. Thus, once we adopt the large-$N_c$
limit and introduce thereby the high radial excitations regarding
them as coupled to a certain local current, we inevitably should
encounter the orbital excitations coupled to the same current,
under the vector mesons we should then understand the mixture
defined above.

\section{Infinite number of states}

The sum rules dealing with infinite number of $V$ and $A$ states
are largely covered in the literature, see
e.g.~\cite{beane,AE,we,peris1,arriola,garda} and references therein.
In this section we give some relevant comments.

First of all, some technical details should be reminded. At large
Euclidean momentum $Q$ the asymptotics of the correlation
functions is~\cite{svz} (see definition~\eqref{trans}),
\be
\label{pert}
\Pi^{V,A}(Q^2)=\frac{N_c}{12\pi^2}\ln\frac{\mu^2}{Q^2}+
\mathcal{O}\left(\frac{\Lambda_{\text{QCD}}^4}{Q^4}\right),
\ee
where we neglect $\mathcal{O}(\alpha_s)$ correction to the
partonic logarithm (the impact of this correction was studied
in~\cite{sc,wepert,pineda}) and the chiral limit is assumed.
In order to obtain
additional constraints on hadron parameters, the conventional
tactics consists in saturating by resonances the individual
correlators $\Pi^{V,A}(Q^2)$ instead of (or along with) the
difference $\Pi^V(Q^2)-\Pi^A(Q^2)$. Evidently, to reproduce the
logarithm in Eq.~\eqref{pert} the infinite number of resonances is
required in the large-$N_c$ limit. Thus, for instance, the vector
correlator can be written as (we omit the complication with the
$D$-wave $V$-states for the reasons explained in Sect.~4.4)
\be
\label{nopert}
\Pi^V(Q^2)=\sum_{n=0}^{\infty}\frac{2F_{V,n}^2}{Q^2+m_{V,n}^2}+\text{S.C.},
\ee
here S.C. means "subtraction constant". It should be noted that
the perturbative continuum is not present any more in
Eq.~\eqref{nopert}, this circumstance expresses the quark-hadron
duality.

The first approximation to the sum in Eq.~\eqref{nopert} is
integral over $n$, this can be explicitly seen via the
Euler-Maclaurin summation formula~\cite{we,peris1},
\begin{multline}
\sum_{n=0}^Nf(n)=\int_0^{N}\!\!f(x)\,dx+\frac12\left[f(0)+f(N)\right]+\\
+\sum_{k=0}^{\infty}
(-1)^k\frac{B_{k+1}}{(2k+2)!}\left[f^{(2k+1)}(N)-f^{(2k+1)}(0)\right],
\label{EM}
\end{multline}
where $B_1=\frac16,\,B_2=\frac{1}{30},\,\dots$ are Bernoulli numbers.
Consequently, the partonic logarithm can be
reproduced if masses and residues are related by
\be
\label{soot}
F_{V,n}^2\sim\frac{dm_{V,n}^2}{dn},
\ee
which should hold at least at $n\rightarrow\infty$. To advance
further one needs some ansatz for $m_{V,n}^2$ or $F_{V,n}^2$. There
is no experimental information on the residues of highly excited
states, but fortunately experiment~\cite{bugg} indicates on the
Regge behaviour of masses, $m_n^2\sim n$. From the theoretical
side, the same behaviour is suggested by quantization of
quasiclassical meson string (see, e.g.,~\cite{sv,mpla,arriola2}).
These
arguments justify the standard use of linear {\it ans\"{a}tze} (up to
corrections vanishing at $n\rightarrow\infty$) in the sum rules
under consideration. Relation~\eqref{soot} gives then constant
residues. To the best of our knowledge, this is the only reason
why one believes that residues are independent of $n$, at least at
large $n$. The question is whether it is possible to justify
qualitatively the constant behaviour of residues independently?

We suggest such argument based on the semiclassical string picture
for mesons. Consider a typical pulsating meson string (see,
e.g.,~\cite{sv} for details), in the large-$N_c$ limit it never
breaks. The mesons decay then due to the electromagnetic
interactions of quark and antiquark --- if they "meet" each other
inside the string and annihilate into the electron-positron pair.
The expectation time $\tau$ for such an event in the pulsating
string is proportional to the string length $l$, the latter is
proportional to the meson mass $m$. Taking into account that
$\tau$ is then typical life-time of the meson, one has
\be
\label{one}
\Gamma_{V\rightarrow
e^+e^-}\sim\frac{1}{\tau}\sim\frac{1}{l}\sim\frac{1}{m}.
\ee
On the other hand, the residues of the vector mesons are related
to their electromagnetic widths as
\be
\label{two}
\Gamma_{V_n\rightarrow e^+e^-}=\frac{4\pi\alpha^2F_{V,n}^2}{3m_{V,n}}\sim
\frac{F_{V,n}^2}{m_{V,n}}.
\ee
Comparing relations~\eqref{one} and~\eqref{two} we conclude that
the residues $F_{V,n}^2$ should not depend on $n$, i.e. they are
constant.

Finally, we arrive at a quite unexpected result: Assuming a
well-motivated string picture for excited mesons, one is able to
argue for the constant residues, then relation~\eqref{soot}, which is a
consequence of quark-hadron duality, yields the linear mass
spectrum, $m_n^2\sim n$, i.e. this linearity may be deduced from
the sum rules without quantization of hadron string.

Another comment concerns derivation of the KSFR relation from the
sum rules. This relation is believed to hold in the phenomenology,
in the sum rules, however, it is usually imposed from outside. As
was argued above, it is natural to expect that $V$ and $A$ meson
residues --- the electromagnetic decay constants --- are a universal
constant, $F_{V,A;n}^2=F^2$. Assuming that this feature extends
up to the $\rho$-meson, the KSFR relation reads $F^2=2f_{\pi}^2$.
In principle, we may choose the slope of the linear mass spectrum
such that the KSFR relation is reproduced due to Eq.~\eqref{soot},
this trick is consistent with the phenomenology~\cite{AE}. An
interesting question rising here is whether it is possible for
the KSFR relation to infer from the sum rules in a way weakly
dependent on a concrete ansatz for the mass spectrum? We will show
that under assumptions above this indeed can be done, the only
assumption about the mass spectrum we need is that going up in
energy one encounters the resonances in the order $V-A-V-A-\dots$\,.
It is a rather weak assumption and consistent with the
phenomenology.

Consider the first Weinberg sum rule~\eqref{w1} with arbitrary
number $N$ of states and take the limit $N\rightarrow\infty$. We
should postulate the pattern of pairing for the $V$ and $A$
states. The most frequent assumptions which one commonly uses are:
(i) the chiral symmetry of QCD provides equal number of $V$ and
$A$ mesons (with reservation on $D$-wave vectors above), so the
$n$-th $A$-meson is paired with the $n$-th $V$-meson; (ii) the
$\rho$-meson is singled out because of the chiral symmetry
breaking at low energies, so the $n$-th $A$-meson is paired with
the $(n+1)$-th $V$-meson.

Consider possibility (i). The sum rule~\eqref{w1} is then
\be
\label{1p}
f_{\pi}^2=\lim_{N\rightarrow\infty}\sum_{n=1}^{N}\left(F_{V,n}^2-F_{A,n}^2\right)=
F^2\sum_{n=1}^{\infty}(-1)^{n+1}.
\ee
Here we encounter a usual problem --- the limit
$N\rightarrow\infty$ leads to ill-defined sums and a generalized
method of summation is needed in order that such sums make a
definite sense. Fortunately, the sum in Eq.~\eqref{1p} is well
known in mathematical analysis, there are many ways of generalized
summation and practically all of them yield the same result
$\frac12$ for the sum in question. Perhaps, the simplest way for
defining this sum is
\be
\label{2p}
\sum_{n=1}^{\infty}(-1)^{n+1}=\lim_{x\rightarrow1-0}\frac{1}{1+x}=\frac12.
\ee
Combining relations~\eqref{1p} and~\eqref{2p} we get the KSFR
relation.

Consider possibility (ii). The sum rule~\eqref{w1} takes the form
\begin{multline}
\label{3p}
f_{\pi}^2=\lim_{N\rightarrow\infty}\left(\sum_{n=1}^{N+1}F_{V,n}^2-\sum_{n=1}^{N}F_{A,n}^2\right)=
F_{V,1}^2-\lim_{N\rightarrow\infty}\sum_{n=1}^{N}\left(F_{A,n}^2-F_{V,n+1}^2\right)\\
=F^2\left(1-\sum_{n=1}^{\infty}(-1)^{n+1}\right),
\end{multline}
where we have used the assumption of $V-A-V-A-\dots$ ordering of
states in masses. Substituting relation~\eqref{2p} to sum rule~\eqref{3p}
we again arrive at the KSFR relation, this universality of result
is quite remarkable.

Since the constant residues imply the linear spectrum with
universal slope,
\be
m^2_{V,A;n}=a(b_{V,A}+n),
\ee
it is instructive to write the first generalized Weinberg sum rule
after the summation over infinite number of states. Applying the
standard summation formula~\eqref{EM} and extracting the
$\mathcal{O}(Q^{-2})$ terms, it reads as follows
\be
\label{sot}
(\text{S.O.T.})_V-F^2\left(b_V-\frac12\right)=
(\text{S.O.T.})_A-F^2\left(b_A-\frac12\right),
\ee
where S.O.T. means "singled out terms". For instance, in
ansatz~\eqref{ans1} and~\eqref{ans2} such terms emerge from the
singled out pion in the $A$-channel and the singled out
$\rho$-meson and the heaviest state $\rho^{(N)}$ in the
$V$-channel, Eq.~\eqref{sot} yields
\be
\label{sot2}
F_{\rho}^2+F_{\rho^{(N)}}^2-F^2\left(b_V-\frac12\right)=
f_{\pi}^2-F^2\left(b_A-\frac12\right),
\ee
Substituting the parameters from ansatz~\eqref{ans1}
and~\eqref{ans2}, we see that Eq.~\eqref{sot2} is satisfied. From
the absence of gauge-invariant condensate of dimension~2~\cite{svz},
leading to the absence of $\mathcal{O}(Q^{-2})$ contribution in
Eq.~\eqref{pert},
the OPE requires that both sides in Eq.~\eqref{sot2} have to be equal
to zero, the considered ansatz meets this condition as well.

The operations with the divergent sums call for another one
comment as they are often present in the sum rules in the
large-$N_c$ limit, sometimes this circumstance causes criticism.
Naively, such operations should indeed lead to very ambiguous
results as long as these sums are ill-defined. This is correct if
one throws away any physical content and considers the matter
mathematically only. The situation then looks as if we worked with
differential equations without boundary conditions, this would be
useless for physics. The role of additional assumptions in the sum
rules is somewhat similar to that of physical boundary conditions.
When imposed correctly, the additional assumptions should always
remove ambiguities.

In this regard, the sum $\sum(-1)^{n+1}$ is a simple educative
example. The answer for this sum depends on a way of grouping the
terms, for this reason the problem is not well defined
mathematically, but in physics the pattern of grouping is fixed by
additional assumption(s). For instance, let us assume that the highly
excited $V$ and $A$ states become exactly degenerate,
$m_{V,n}=m_{A,n}$, $F_{V,n}=F_{A,n}$ at $n\geq N$. This means that
going up in energy, since the scale $m_{V,N}$ the $V$ and $A$ resonances
will be coming in pairs. This feature provides the {\it physical
pattern} of grouping the individual contributions in that part of
the spectrum,
\be
1-1+1-1+\dots=(1-1)+(1-1)+\dots=0+0+\dots=0,
\ee
the same pattern will hold in the higher-order sum rules. Such
a local conspiracy results in cancellations of the same type as
cancellation of $V$ and $A$ perturbative continuums in the
difference $\Pi^V(Q^2)-\Pi^A(Q^2)$, in this sense one may think of
a certain duality between that part of spectrum and perturbative
continuum. It should be noted incidentally that assuming some pattern of
pairing of states, any regularization of divergent sums has to respect
this pattern, otherwise result will be senseless.

Let us assume now a more realistic case --- the $V$ and $A$ mesons
are not exactly degenerate at any finite $N$. This means that
going up in energy we will be encountering the resonances
consecutively state by state, hence, the summation has to be
performed in the same way. For validity of this statement it is
not necessary to require the $V-A-V-A-\dots$ recurrence, if the number
of $V$ and $A$ contributions is almost equal (say, if the $V$ and
$A$ mesons form approximately degenerate parity doublets), the
permutation of some $V$ and $A$ contributions does not change the
result. The partial sums of $\sum(-1)^{n+1}$ will be then either
$0$ or $1$. The generalized summations yield the averaged value
$\frac12$. This result is in one-to-one correspondence with the
fact that the Euclidean behaviour of correlators is sensitive only
to averaged features of the spectral densities, for this reason
the generalized summations are usually quite effective tools in
the Euclidean domain.

\section{Conclusions}

We have considered various aspects of generalized Weinberg sum
rules at different saturation schemes. Our main conclusions can be
summarized as follows.

\begin{itemize}
  \item When one imposes the asymptotic chiral symmetry on
  phenomenological Lagrangians, the Weinberg like sum rules follow
  naturally. The difficulties related to the operator product
  expansion for correlation functions are problems of a specific
  derivation of the sum rules rather than problems of the sum
  rules themselves. In principle, one may use such sum rules
  including up to infinite number of narrow resonances having
  forgotten about those problems.
  \item The simplest ansatz for saturation of the Weinberg like sum
  rules compatible with the phenomenology is not the standard
  $\pi-\rho-a_1$ one, but rather the $\pi-\rho$ scheme. The
  simplest possibility for a nontrivial parity doubling appears at
  the $\pi-\rho-a_1-\rho(1450)$ saturation.
  \item The string like picture for mesons suggests that the meson
  decay constants are independent of the number of
  excitation. The quark-hadron duality requires then the linear
  spectrum in the large-$N_c$ limit, i.e. the spectrum of typical
  quantized string. This observation provides an additional
  argument in favour of old hypothesis that QCD in the large-$N_c$
  limit is dual to some string theory.
\end{itemize}

Finally, we have tried to demonstrate that the potential of
considered QCD sum rules is not exhausted, further investigations
in this direction may lead to interesting results and
applications. Hopefully, the present revision of these known sum
rules will be useful in this respect.

\section*{Acknowledgments}

The work was supported by the grants RFBR 05-02-17477,
LSS-5538.2006.2, and by the
Ministry of Education of Russian Federation, grant RNP.2.1.1.1112.

\end{document}